# Volcano-Like Ferroic Transitions Deviating from the Model of Landau Theory

Yuxuan Sheng and Menghao Wu[*]

School of Physics and School of Chemistry, Huazhong University of Science and Technology, Wuhan, Hubei 430074, China

Abstract   We predict the existence of abnormal volcano-like temperature dependence of polarization or magnetization with maxima located at elevated temperature, distinct from classical model based on Landau theory. One case is ferroelectricity with long ion displacements and quantized polarizations that cannot be used for expansion in Landau model, and the switching pathway involves various metastable phases where the polar phase is higher both in energy and entropy compared with non-polar phase. Another case is compensated antiferromagnets with two opposite spin lattices of different spin exchange constants. Such difference can be utilized for a unique type of "temperature differentiated multiferroicity", where large magnetizations can be reversed upon ferroelectric switching between two Curie temperature with alternating half of spins in paramagnetic state. We demonstrate these proposals by first-principles calculations on several paradigmatic systems, including magnetic bilayers intercalated by Ag ions or metal molecules.

Ferroic materials are characterized by the presence of a spontaneous and switchable order parameter arising from spontaneous symmetry breaking below Curie temperature ($T_c$), such as polarization in ferroelectrics and magnetization in ferromagnets. The ferroic phase transition can be described by Landau theory based on an expansion of the free energy $F$ of a thermodynamic system[1] in terms of an order parameter $\chi$:

$$F(T) = F_0 + \frac{1}{2}\alpha(T - T_c)\chi^2 + \frac{1}{4}\beta\chi^4$$

where $\beta > 0$ ensures stability. When $T < T_c$, such formula give rises to a typical double-well potential for bi-stable states. The equilibrium configuration is determined by the minima of $F$:

$$\frac{\partial F}{\partial \chi} = \alpha(T - T_c)\chi + \beta\chi^3 = 0$$

$$\chi = \begin{cases} 0, & T > T_c \\ \pm\sqrt{\frac{\alpha(T_c - T)}{\beta}}, & T < T_c \end{cases}$$

indicating a continuous emergence of long-range ferroic order through a second-order phase transition as thermal fluctuations subside. As the temperature rises and exceeds $T_c$, $\alpha(T_c - T)$ become negative, driving the order parameter to zero and leading to the emergence of the disorder para-phase. Such Landau phenomenological model has been successfully applied to various ferroic systems in the past century, also including emerging new types including sliding ferroelectricity,[2] altermagnetism[3] and autferroicity[4].

In this paper, we predict some cases that deviate from classical model of ferroic transition, where the ferroelectric polarization or magnetization diminishes at 0K while reaching the maxima at elevated temperature. Compared with typical ferroelectricity with small deviations from paraelectric phase as transition state, some unconventional properties of ferroelectricity with long ion displacements have been explored recently, including quantized[5,6] or fractional quantum ferroelectricity[7] where ferroelectric switching involves ion migrations for multiple or fractional lattice constants, and such ferroelectricity may even exist in non-polar crystals.[8] Here we propose that the Landau theory of phase transitions is not applicable in such ferroelectricity where the large quantized polarization cannot be used to expand as a small amount. Similar anomalous temperature dependence also emerges in some magnetic systems including compensated antiferromagnets, which becomes uncompensated at elevated temperature due to two spin lattice with different exchange constants. We further show first-principle evidence that such uncompensated magnetization can be reversed via ferroelectric switching in some systems, which is long-sought but elusive in current single-phase multiferroics, noting that type-I multiferroics are mostly with negligible magnetoelectric couplings while type-II multiferroics possess much weaker polarizations and magnetizations.[9,10]

We first propose a possible factor that may induce such atypical transition in ferroelectricity. The ferroelectric switching pathway of long ion displacements may involve various metastable phases, e.g., $CuCrX_2$ (X=S, Se) with Cu in hex-coordination as the metastable paraelectric phase compared with the ferroelectric ground state with Cu in tetra-coordination.[11] Their ferroelectricity has been experimentally confirmed in recent reports,[12-14] including the quantized in-plane ferroelectricity violating Neumann's principle, and similar ferroelectricity has also been detected in thin-layer $AgCrS_2$.[15,16] The volcano-like behavior may take place if the polar phase possesses higher entropy $S$ compared with non-polar phase that is slight lower in energy, so the polar phase will dominate with lower $F=E-TS$ above critical temperature, as shown in Fig. 1(b). Herein $CuCrX_2$ is not a candidate since the ferroelectric phase is the ground state. However, bilayer $CrS_2$ intercalated by Ag (i.e., $AgCr_2S_4$)[17,18] ions that are inclined to form di-coordinate configurations (denoted as DI phase) are non-polar,[19] while the tetra-coordinate configurations are ferroelectric (denoted as FE phase) that is slightly higher in both energy and entropy (see Fig. 2(a)). Compared with the ground state DI phase, FE phase possesses smaller interlayer distance $d$ and its intercalated Ag layer is vertically displaced from the middle plane. Also distinct from the typical double-well model with a transition state as the paraelectric (PE) state, as shown in Fig. 2(b), in the ferroelectric switching pathway where Ag ions are displaced by 1/3 lattice constant horizontally, the PE state is metastable so soft modes vital in the formation of conventional ferroelectricity are absent here. Our *ab initio* molecular dynamics (AIMD) simulations in Fig .2(c) indicate the phase transition in 400-600K, which can also be revealed by the evolution of average interlayer distance $d$ in Fig 2(d) that drastically shrinks from ~4.6 to ~3.6 Å at 600 K. The average vertical displacement of Ag ions $\Delta d = d_1 - d_2$ (marked in Fig.

2(c)) also increases to above 0.4 Å at 600K from close to zero at 400K, revealing the occurrence of paraelectric-ferroelectric transition at elevated temperature.

Machine-learning-based force field is further used for large-scale deep potential molecular dynamics (DPMD) simulations of phase transition (see the method in Supplemental Material)[20]. The trained potential field for calculating the force and energy seems to approach the accuracy of AIMD, as shown in Fig. 3(a), where the root mean square error (RMSE) for the force and energy are respectively estimated to be ~0.046 eV/Å and ~1.13 meV/atom after systematic training, and the energy evolution for FE-DI transition is also well reproduced in Fig. 3(b). Herein a 252-atom supercell was adopted in DPMD simulations for 120 ps relaxation at each given temperature, where a step-like drastic decline of the average interlayer distance $d$ is observed at ~470 K, indicating the occurrence of DI-FE phase transition. The average vertical displacement of Ag ions $\Delta d = d_1 - d_2$, which is closely related to the macroscopic polarization, also increase from close to zero at ~420K, to its maxima at ~580 K.

We further propose another possible factor that may induce similar volcano-like transition in magnetic systems. Suppose there is a ferrimagnetic system composed of two opposite spin lattices, with distinct spin exchange constants that lead to distinct Curie temperature $T_1$ and $T_2$ ($T_2 > T_1$), as shown in Fig. 4(a). If the coupling between them is negligible, according to the Landau theory, the net magnetization of two layers $M = M_1 - M_2$ may increase with $T$ when $T < T_1$. In particular, this can be applicable to compensated antiferromagnets (sometimes denoted as Luttinger compensated magnets), which may become uncompensated at elevated temperature, especially when one of the spin lattices becomes paramagnetic leaving the other ferromagnetic. Such candidates include various magnetic bilayer sliding ferroelectrics where the interlayer coupling is antiferromagnetic and the intralayer coupling is ferromagnetic.[21] The Curie temperature of the upper and down layer can be slightly different due to the interlayer inequivalency induced by vertical polarizations. Similar difference in Curie temperature induced by electrical polarization may also emerge when a homobilayer ferromagnet is separated by a ferroelectric monolayer that also induces antiferromagnetic interlayer coupling, as illustrated in Fig. 4(b): the bilayer system with one ferromagnetic layer and the other close to paraelectric phase between $T_1$ and $T_2$ exhibits a non-zero magnetization, which can be reversed upon ferroelectric switching due to swapping of Curie temperature between two layers.

Such atypical multiferroic coupling, denoted as "temperature differentiated multiferroicity", may render much larger switchable magnetization compared with either type-I or type-II multiferroics in previous studies. Here we select a paradigmatic case based on our previous studies[22,23]: when magnetic bilayers (e.g., $VSe_2$, $CrI_3$) are intercalated by molecules like titanium porphyrin (TiP) or vanadium porphyrin (VP), each metal ion will be inclined to form a bond with one of the layer, which gives rise to a vertical polarization switchable upon sliding of molecules and vertical displacements of the metal ions. In such bilayer with intralayer ferromagnetism and interlayer antiferromagnetic couplings, a considerable difference in the Curie temperature of two layers can be induced by the vertical polarization. Our predictions can be demonstrated by Monte Carlo simulations based on Heisenberg model:

$$\widehat{H} = -\frac{1}{2}\sum_{\langle ij \rangle} J\vec{S}_i \cdot \vec{S}_j - \sum_{\langle i \rangle} D(S_{iz})^2$$

where $J$ is defined as the nearest neighboring exchange coupling parameter, $D$ (estimated to be 2.16 and 1.20 meV for monolayer $VSe_2$ and $CrI_3$) is the single-site magnetic anisotropy energy, and $|S|=1/2$ and $3/2$ respectively for monolayer $VSe_2$ and $CrI_3$. The calculated $J$ for the $VSe_2$ layer binding with TiP molecules and pristine $VSe_2$ monolayer are 160.63 meV and 109.38 meV, respectively. Such difference is also notable in $CrI_3$ layer binding with VP molecules and pristine $CrI_3$ with $J$ estimated to be respectively 5.6 and 2.8 meV. Upon the large interlayer difference in magnetic exchange parameters for intercalated bilayer $VSe_2$ and $CrI_3$, together with negligible interlayer coupling, large switchable net magnetizations are formed and reach the maximum values of respectively 10.74 and 19.95 $\mu_B$ per unitcell (respectively containing 32 V ions and 16 Cr ions) at $T = T_1$, as shown in Fig. 5(a) and (b).

In summary, we predict some exceptional cases with volcano-type temperature dependence where the polarization or magnetization increases with temperature from zero to the peak. Ferroelectric transition described by Landau theory is based on an expansion of the free energy in terms of polarization, while the large quantized polarization of unconventional ferroelectricity with long ion displacement cannot be used for expansion of free energy as a small amount. The switching pathway of long ion displacements may involve various metastable phases, also distinct from the long-established double-well potential model and soft mode model, and the volcano-like behavior may emerge if the polar phase is higher both in energy and entropy compared with non-polar phase. The volcano-like temperature dependence of magnetization may also take place in compensated antiferromagnets with two opposite spin lattices of different spin exchange parameters. Such difference stems from polarizations in some multiferroic systems, where ferroelectric switching swaps the paramagnetic state and ferromagnetic state between two opposite spin lattices and reverses large magnetization, distinct from multiferroic couplings in previous reports. We demonstrate these proposals by first-principles calculations assisted by machine-learning-based force field and Monte Carlo simulations on several paradigmatic systems, including magnetic bilayers intercalated by Ag ions or metal molecules.

Acknowledgements--

We thank Prof. Jun-Ming Liu for helpful discussion.

Notes

The authors declare no conflict of interest.

Supplemental Material

The supporting information includes the computational methods.

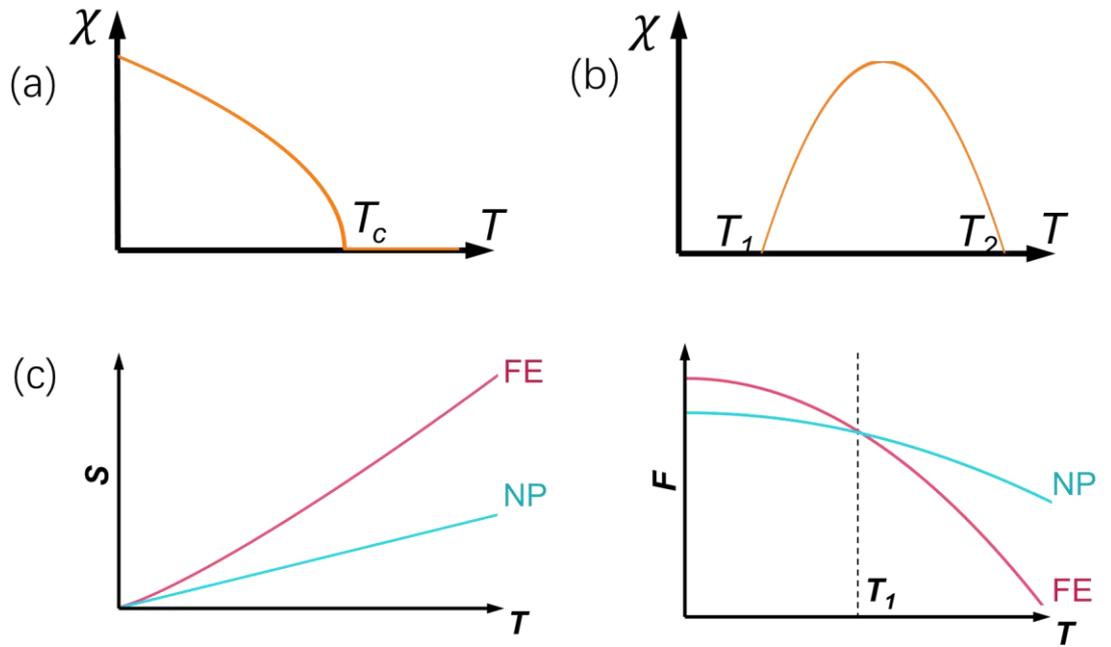

FIG 1. Comparison between (a) Landau theory description of ferroic transition and (b) our predicted volcano-like temperature dependence. (c) The temperature dependence of entropy and free energy for ferroelectric (FE) and non-polar (NP) phase in entropy-driven ferroelectric transition, where FE phase is slightly higher in energy and entropy compared with NP phase, and FE phase will become more favorable in free energy above critical temperature.

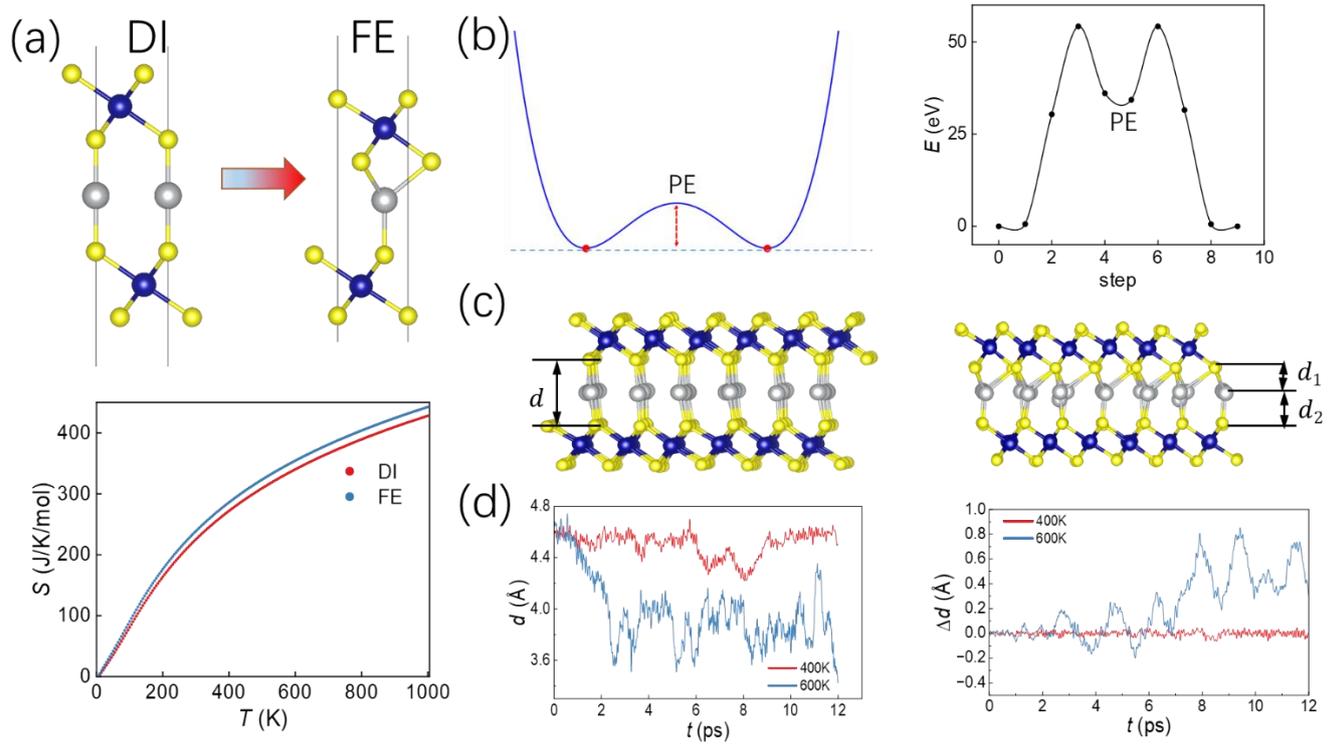

FIG. 2 (a) DI and FE phase of AgCr$_2$S$_4$, and comparison of their entropy. (b) Double-well potential of classical ferroelectricity with PE state as a transition state, compared with the switching pathway of AgCr$_2$S$_4$ with PE state as a metastable state. (c) Snapshots of the equilibrium structures for AgCr$_2$S$_4$ at 400K and 600K at the end of 12 ps in AIMD simulations, and (d) corresponding evolution of interlayer distance $d$ and Ag ion displacement $\Delta d = d_1 - d_2$.

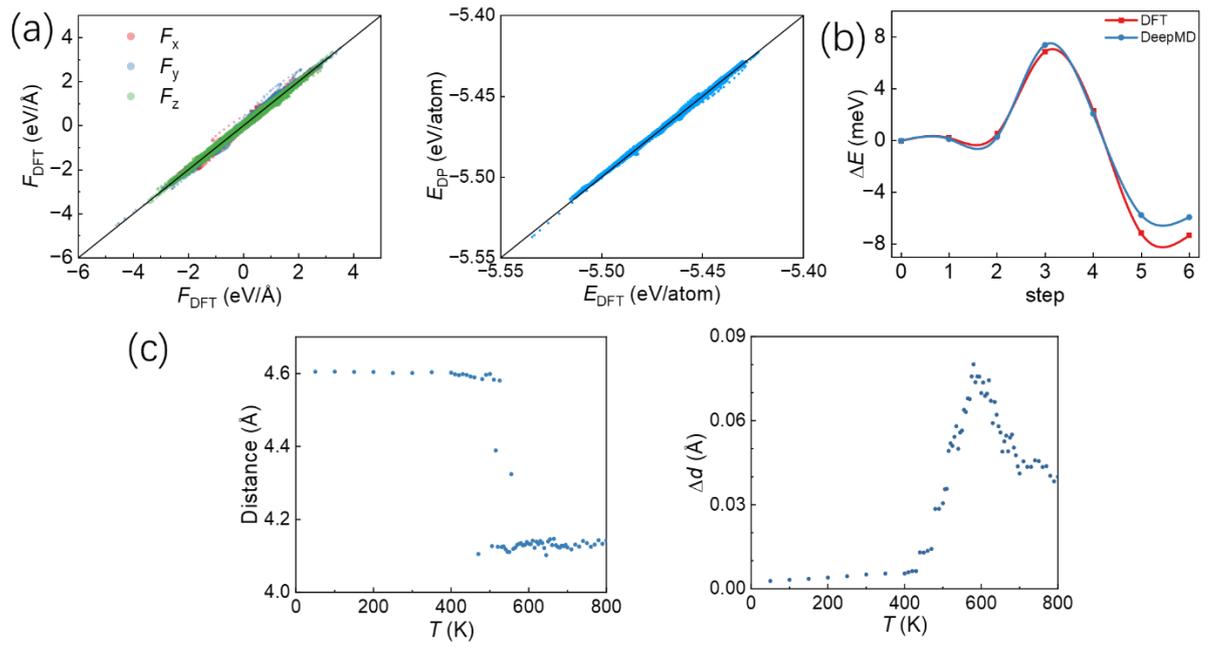

FIG. 3 Comparison of (a) atomic forces, energies and (b) FE-DI transition pathway of $AgCr_2S_4$ calculated by DP model and DFT. The temperature dependence of (c) interlayer distance $d$ and Ag ion displacement $\Delta d = d_1 - d_2$.

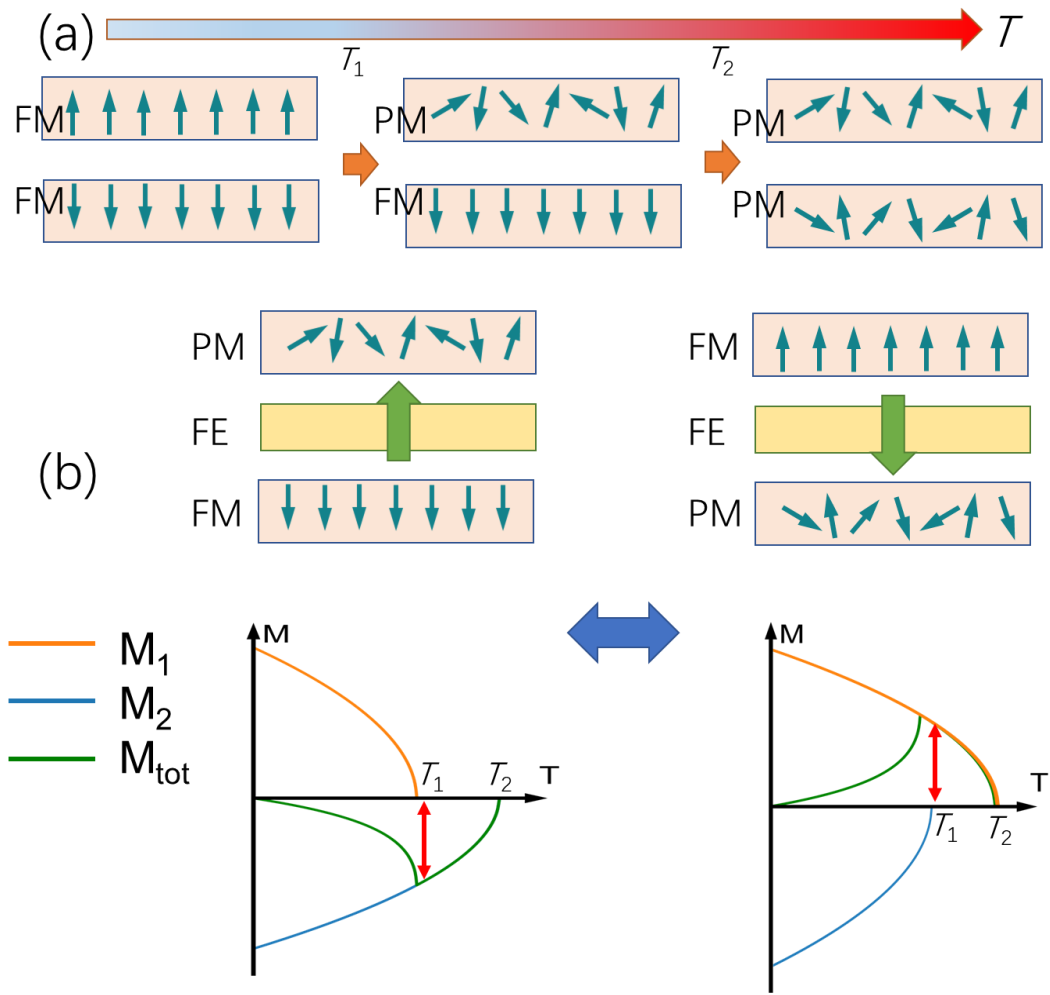

FIG. 4 (a) Evolution of two ferromagnetic (FM) layers/lattices with different Curie temperature that are antiferromagnetically coupled, where the compensated magnetization at 0K becomes uncompensated as only one of them become paramagnetic (PM) between two Curie temperature. (b) A sketch of temperature differentiated multiferroicity: when the difference in Curie temperature between the top and bottom layer is induced by polarization of ferroelectric (FE) middle layer, a considerable uncompensated magnetization $M_{tot}=M_1+M_2$ marked by red arrow can be formed, which can be reversed via ferroelectric switching. The spin and polarization directions are respectively marked by olive and green arrows.

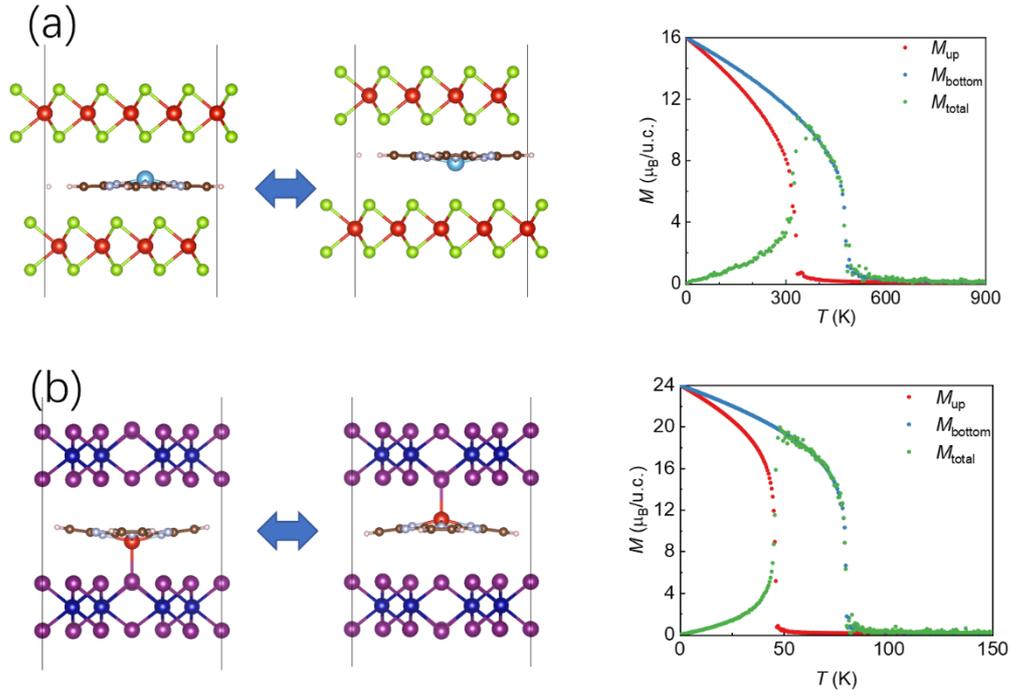

FIG. 5 Ferroelectric switching between bi-stable states and temperature dependence of magnetization for (a) bilayer $VSe_2$ intercalated by TiP molecules and (b) bilayer $CrI_3$ intercalated by VP molecules. The magnetizations per unitcell (u.c.) of upper layer, bottom layer and the whole bilayer are respectively marked by red, blue, and green dot lines.